\documentclass[journal=jacsat,manuscript=article]{achemso}

\usepackage{chemformula} 
\usepackage[T1]{fontenc} 

\author{Xi Yang}
\affiliation[First University]
{Frontiers Science Center for Nano-optoelectronics and State Key Laboratory for Mesoscopic Physics, School of Physics, Peking University, Beijing 100871, China}
\author{Shui-Jing Tang}
\affiliation[First University]
{Frontiers Science Center for Nano-optoelectronics and State Key Laboratory for Mesoscopic Physics, School of Physics, Peking University, Beijing 100871, China}
\author{Jia-Wei Meng}
\affiliation[First University]
{Frontiers Science Center for Nano-optoelectronics and State Key Laboratory for Mesoscopic Physics, School of Physics, Peking University, Beijing 100871, China}
\author{{Pei-Ji Zhang}}
\affiliation[First University]
{Frontiers Science Center for Nano-optoelectronics and State Key Laboratory for Mesoscopic Physics, School of Physics, Peking University, Beijing 100871, China}
\author{You-Ling Chen}
\email{ylchen@semi.ac.cn}
\affiliation[First University]
{State Key Laboratory on Integrated Optoelectronics, Institute of
Semiconductors, Chinese Academy of Sciences, Beijing 100083, China}
\author{Yun-Feng Xiao}
\email{yfxiao@pku.edu.cn}
\affiliation[First University]
{Frontiers Science Center for Nano-optoelectronics and State Key Laboratory for Mesoscopic Physics, School of Physics, Peking University, Beijing 100871, China}
\alsoaffiliation[Second University]
{Collaborative Innovation Center of Extreme Optics, Shanxi University, Taiyuan 030006, China}
\alsoaffiliation[Third University]
{Peking University Yangtze Delta Institute of Optoelectronics, Nantong 226010, China}
\alsoaffiliation[Fourth University]
{National Biomedical Imaging Center, Peking University, Beijing 100871, China}

\title[An \textsf{achemso} demo]
  {Phase-transition microcavity laser}

\abbreviations{IR,NMR,UV}

\keywords{Microcavity laser, phase transition, liquid crystal, thermal sensing}

\begin{document}


\begin{abstract} 

Liquid-crystal microcavity lasers have attracted considerable attention because of their extraordinary tunability and sensitive response to external stimuli, and they operate generally within a specific phase.
Here, we demonstrate a liquid-crystal microcavity laser operated in the phase transition, 
in which the reorientation of liquid-crystal molecules occurs from aligned to disordered states.
A significant wavelength shift of the microlaser is observed, resulting from the dramatic changes in the refractive index of liquid-crystal microdroplets during the phase transition. 
This phase-transition microcavity laser is then exploited for sensitive thermal sensing, enabling two-order-of-magnitude enhancement in sensitivity compared with the nematic-phase microlaser operated far from the transition point. Experimentally, we demonstrate an exceptional sensitivity of -40 nm/K and an ultrahigh resolution of 320 \textmu K. The phase-transition microcavity laser features compactness, softness, and tunability, showing great potential for high-performance sensors, optical modulators, and soft matter photonics.
\end{abstract}

\section{Introduction}

Optical microcavities with high quality factors and small mode volumes  
significantly enhance the light-matter interaction, 
facilitating a wide variety of applications in nonlinear optics,\cite{lai2020earth,li2018whispering,strekalov2016nonlinear,del2007optical}
quantum optomechanics,\cite{aspelmeyer2014cavity,zhang2021optomechanical} non-Hermitian physics,\cite{ozdemir2019parity,cao2015dielectric}
as well as high-performance lasers and sensors.\cite{toropov2021review,liu2020nonlinear,basiri2019precision} 
Generally, the emergence of new materials may improve performance and expand many functionalities of optical microcavities.\cite{liu2022emerging} For example, lithium niobate with a large electro-optic coefficient and a second-order nonlinear coefficient, greatly promotes the development of microcavity-based  electro-optic modulators,\cite{wang2018nanophotonic} frequency combs,\cite{zhang2019broadband,wang20222} and nonlinear frequency converters.\cite{zhu2021integrated,lu2020toward,lin2019broadband,zheng2019high}
Halide perovskite featuring the intrinsic gain inspires the integrated microcavity lasers toward applications of all-optical switching,\cite{huang2020ultrafast} color display,\cite{gao2018lead} and light-emitting diodes. \cite{miao2020microcavity,zhang2021halide}

Liquid crystal (LC) has been recently introduced into optical microcavities, creating new opportunities in tunable lasers and high-performance sensors.\cite{mysliwiec2021liquid,coles2010liquid,humar2016liquid,ma2022self}
The long-range order and the fluid nature of LC molecules lead to its unique properties not being presented in any other material.
The exceptional tunability enables a wide range of applications of LC microlasers in tunable light sources,\cite{xiang2016electrically,chen2022over,adamow2020electrically} light field modulation,\cite{zhang2020tunable,papivc2021topological,ali2022demand} and laser display.\cite{zhan20213d,wang2021programmable} The stimulus-response allows LC microlasers for high-performance sensing of temperature\cite{franklin2021bioresorbable,humar2017biomaterial}  and   humidity\cite{hu2020snr}, as well as  biomolecules\cite{gong2021topological,wang2021applications} and ions\cite{duan2019detection,wang2018detecting}. In these applications, the LC microlasers are tuned generally by changing the orientation of LC molecules within a specific phase. 
Theoretically, LC microlasers operated in the vicinity of phase transition are expected to enhance significantly their tunability and sensitivity,\cite{franklin2021bioresorbable,yang2021operando} but yet to be demonstrated experimentally.

Here we demonstrate a phase-transition microcavity laser using LC microdroplets supporting whispering-gallery modes (WGMs). 
It is observed experimentally that the lasing wavelength experiences a dramatic shift, which reaches the maximum when the LC molecules approach the transition point between two different phases.
This phase-transition microlaser is further utilized for thermal sensing.
The thermal sensitivity in the proximity of phase transition is enhanced by two orders of magnitude without significant changes in background fluctuations, achieving a peak sensitivity of -40 nm/K and an ultrahigh resolution of 320 \textmu K. This sensor featuring phase transition is also compatible with sensitivity-enhanced techniques such as optical field confinement and nonlinear enhancement.\cite{jiang2020whispering,liu2020nonlinear}

\section{Results and discussion}

The schematic diagram of the phase-transition microcavity laser  
is illustrated in Figure 1, in which a LC microdroplet is formed in an aqueous environment due to its hydrophobic properties. 
The rod-like LC molecules are self-assembled with their long axis aligned along the radial direction of the microdroplet by confining surfaces.\cite{humar2009electrically} 
When the LC microdroplet doped with laser dye is pumped by a pulsed laser, multiple WGMs are excited to achieve laser emission (Figure 1a). 
The lasing spectrum depends highly on the LC phases with the specific spatial ordering of molecules, which can be controlled by temperature (Figures 1b and 1c).\cite{yang2014fundamentals}
(i) In the nematic phase, the radially aligned LC molecules 
slightly tilt with increasing temperature, causing a small blueshift of lasing wavelength due to the negative thermo-optic coefficient.
(ii) In the isotropic phase, the LC molecules have no orientational order and can point in any direction, in which the lasing wavelength shows no evident changes.
(iii) In the vicinity of the transition point between two phases, the LC molecules convert from aligned to disordered states, which induces a dramatic change in the refractive index of the LC microdroplet and thus a sharp shift of lasing wavelength.

\captionsetup[figure]{labelfont={bf}, name={Figure}, labelsep=period}
\begin{figure*}[htbp]
\centering
\includegraphics[width=12cm]{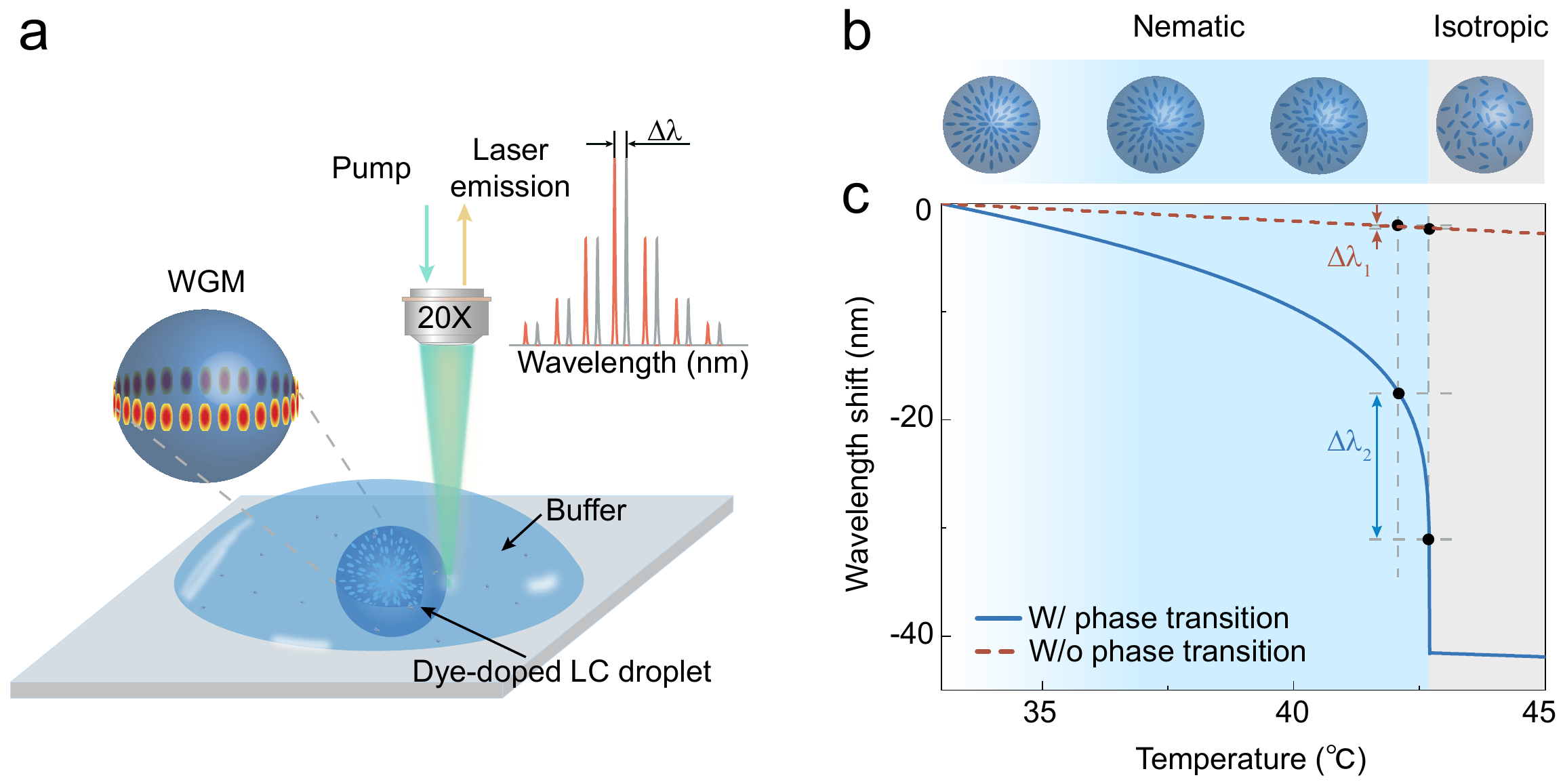}
\caption{\label{fig1}
Concept of the phase-transition microcavity laser. (a) Schematics of the characterization system. The LC microdroplet supporting WGMs is pumped by a pulsed laser, and its lasing spectra are collected by an objective. Inset: thermal-induced spectrum variation of the phase-transition microlaser. (b) Spatial ordering of LC molecules from nematic to isotropic phases.
(c) Dependence of the lasing wavelength on temperature with (blue solid curve) and without (red dashed line) the LC phase transition. 
}
\end{figure*}

Experimentally, liquid crystal 7CB molecules are self-assembled into dye-doped microdroplet  cavities with a diameter of about 15 \textmu m (see Methods in Supporting Information). The surface of the microdroplet is modified by sodium dodecyl sulfate (SDS) to form the radial configuration of LC molecules. The concentration of SDS should be high enough to keep the radial configuration of LC molecules and also be lower to minimize the influence on phase transition. Their spatial ordering inside the microdroplet can be observed from the bright-field and polarized optical images (insets of Figures 2a, 2b, and S1). Typically, a nematic LC microdroplet indicates a cross pattern in the polarized optical image and creates a point topological defect in the center of the droplet under bright-field image (insets of Figure 2a), while these features disappear in the isotropic phase (insets of Figure 2b). 

To characterize the phase transition, the LC microdroplet is pumped by a pulsed laser (532 nm, 1.8 ns) to generate lasing at around 615 nm.
The lasing spectra at the nematic and isotropic phases are presented in Figures 2a and 2b, respectively. It is observed that the transverse-magnetic (TM) lasing modes dominate in the nematic phase, since these modes with electric field oscillating along the long axis of the LC molecules experience a larger dielectric constant and thus higher quality factors than the transverse-electric (TE) modes whose electric field oscillate along the short axis of LC molecules. By contrast, both TM and TE modes lase in the isotropic phase due to the vanish of birefringence (Figure S2).  
It is also found that both lasing threshold and free spectral ranges (FSRs) vary significantly with the phase transition as shown in Figures 2c and 2d, respectively. 
The lasing threshold achieved in nematic LC microlaser (8 \textmu J/mm$^{2}$) is much lower than that in the isotropic phase (110 \textmu J/mm$^{2}$), which is attributed to the phase-transition-induced decrease of refractive index (Figure S3).
The FSR increases statistically from 3.7 THz to 4.1 THz, in which the FSRs of the same mode family (lasing modes with same radial and azimuthal mode numbers, as well as different angular mode number) of a LC microlaser are measured (Figure 2d).

\begin{figure*}[htbp]
\centering
\includegraphics[width=12cm]{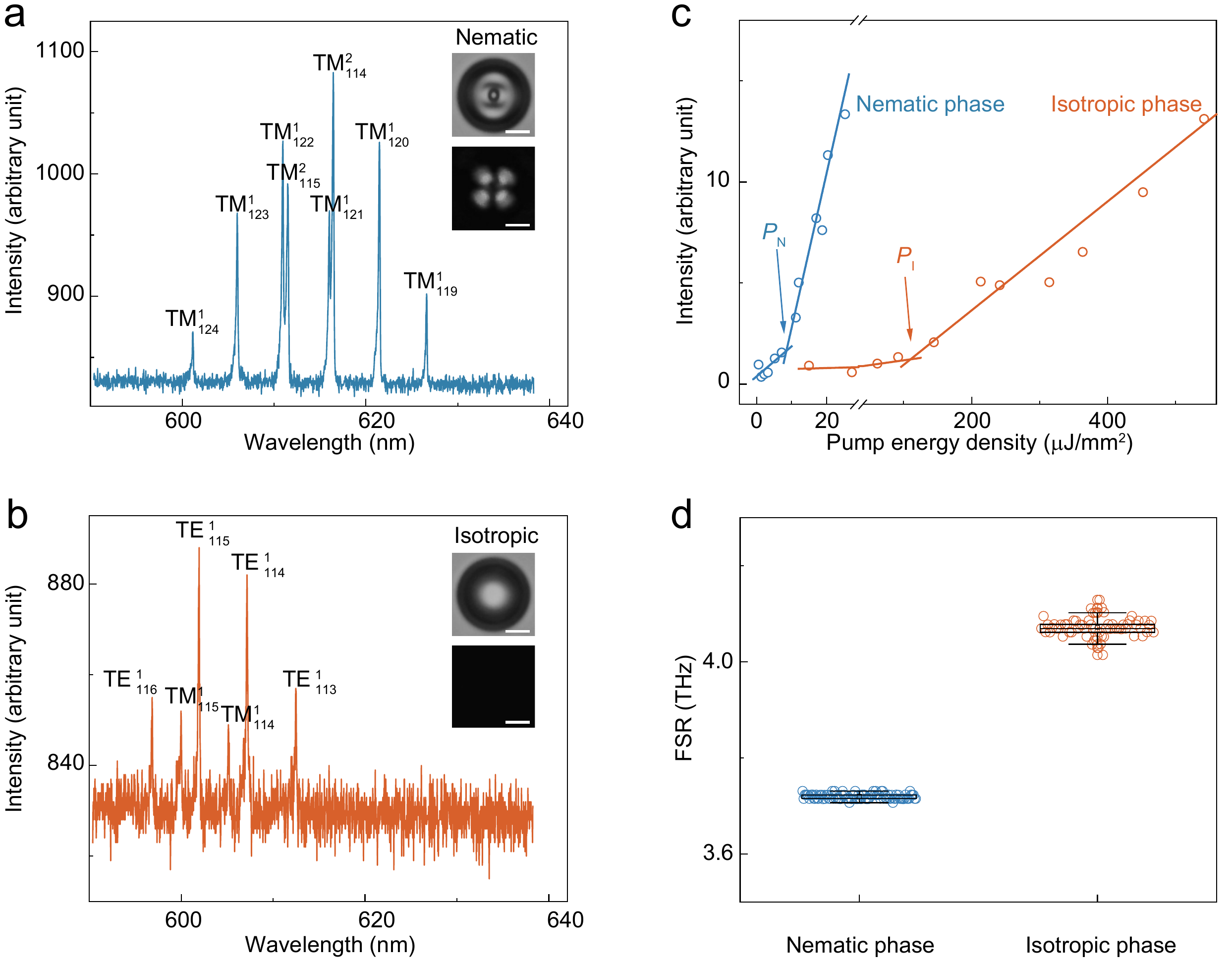}
\caption{\label{fig2}
Characterization of phase-transition microcavity laser. Typical lasing spectra of LC microdroplet with nematic (a) and isotropic phase (b). 
The radial ($q$) and angular ($l$) mode numbers are labeled with $\text{TM}^{q}_{l}$ and $\text{TE}^{q}_{l}$.\cite{gorodetsky2006geometrical} 
The insets show the corresponding  bright-field images (top) and polarized optical images (bottom). Scale bars, 5 \textmu m. (c) Threshold curves of nematic and isotropic LC microdroplets. (d) Statistics of the FSR from the same LC microdroplet in the nematic and isotropic phases.
}
\end{figure*}

The phase-transition microlaser is then exploited for thermal sensing. To explore the sensing performance at different phases, the temperature around the microdroplet is tuned precisely by the stage-top incubator with a proportional-integral-derivative (PID) controller. 
The lasing spectra are recorded and shown in Figure 3a, which depend highly on the environment temperature.  
It is found that the lasing wavelengths show slight blueshifts with increasing temperature, and then experience a sharp variation when approaching the phase transition point (42.7 $^{\circ}$C). The transition process from nematic to isotropic phase is observed by the lasing behaviour of microcavity.
This unique property promises an ultrahigh sensitivity for thermal measurement.      

A four-parameter model is employed to analyze this phase-transition behavior. 
The microlaser with TM modes features electric field oscillating along the long axis of the LC molecules, which corresponds to the extraordinary refractive index (${n_e}$) of the nematic LC microdroplet.
This refractive index ${n_e}$ depends highly on the temperature and can be expressed as (See Supporting Information)\cite{li2004temperature}

\begin{equation}
    {n_e}(T)=A-BT+\frac{2(\Delta{n})_0}{3}(1-\frac{T}{T_C})^\beta
\end{equation}
where ${A}$ and ${B}$ are the material constants that can be obtained by fitting the temperature-dependent average refractive index $<n>=(n_e+2n_o)/3$ ($n_o$ is the ordinary refractive index). $(\Delta{n})_0$ and ${\beta}$ are the birefringence at ${T}$=0 and material constant, respectively, which can be obtained from fitting the temperature-dependent birefringence $\Delta{n}$=${n_e}$-${n_o}$. ${T_C}$ is the phase-transition temperature (i.e., cleaning point).
It indicates that refractive index ${n_e}$ decreases with increasing temperature ${T}$, and the variation of ${n_e}$ is dominant by the last term when the LC microlaser is operated in the vicinity of the phase transition.
This enables a negative thermo-optic coefficient, which reaches the maximum near the phase transition (Figure S2). 
On the other hand, birefringence vanishes in the isotropic phase (i.e., $n_e=n_o$), and the refractive index decreases linearly with temperature.  
Therefore, the dependence of resonant wavelengths on temperature is calculated and shown in Figure 3b, in good agreement with the experimental results in Figure 3a.

To evaluate the sensitivity of the microlaser-based thermal sensor, the lasing wavelength is extracted from the spectrum by Gaussian fitting the experimental results.\cite{schubert2020monitoring} 
The dependence of lasing wavelength on temperature is shown in Figure 3c (blue circles). It is found that the lasing wavelength blueshifts faster and reaches the maximum of -18 nm when the temperature approaches the transition point.  
The sensitivity ${S}=-\Delta\lambda/\Delta{T}$ is shown in Figure 3c (orange circles). In the nematic phase, the sensitivity is only -0.4 nm/K, while it boosts to -40 nm/K in the vicinity of phase transition.
It shows two-order-of-magnitude improvement in sensitivity, which is ascribed to the reorientation of LC molecules from aligned to disordered states during the phase transition.
Moreover, the temporal response of this thermal sensor is studied by changing the temperature repeatedly (Figure 3d, grey curve).  
It is found that the evolution of lasing wavelength follows instantaneously the environmental temperature. 
Note that the phase-transition temperature of the microlaser can be tuned by tailoring different LC materials to satisfy the application requirement.\cite{van2012combinatorial}
The ultrahigh sensitivity in the vicinity of phase transition and the real-time response of LC microlaser make it quite suitable for monitoring the subtle fluctuation at a specific temperature.

\begin{figure*}[htbp]
\centering
\includegraphics[width=12cm]{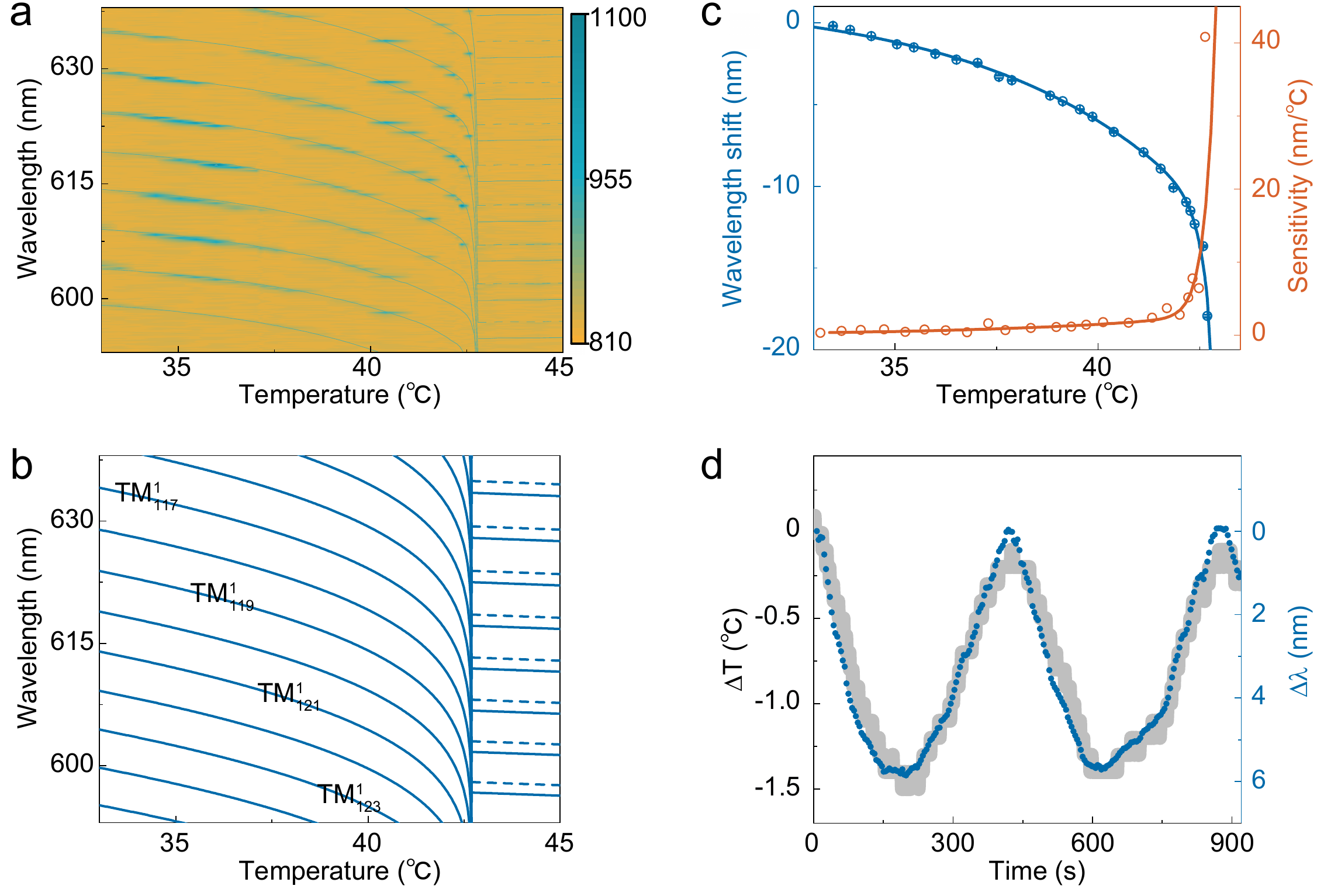}
\caption{\label{fig3}
Phase-transition-enhanced thermal sensing. (a) Experimental lasing spectra of LC microlaser at different temperatures. (b) Theoretical resonant wavelengths at different temperatures obtained from four-parameter model\cite{li2004temperature}. (c) Wavelength shift of a specific lasing mode (blue circles) and the corresponding sensitivity in thermal sensing (orange circles) extracted from a.
The error bar is obtained from five measurements. Solid curves, fitting results.  
(d) Temporal response of the lasing wavelength when operated near the phase-transition point (blue dots). The real-time temperature is indicated with the grey curve. 
}
\end{figure*}

To analyze the resolution for thermal sensing, the stability of the lasing wavelengths is studied. Its lasing spectra are monitored for 5 minutes at different temperatures, which are tuned from the nematic phase to approach the phase transition (Figure S4).
The fluctuations of lasing wavelengths in three representative temperatures are presented in Figure 4a. At temperatures of 32.8 $^\circ$C, 41.9 $^\circ$C, and 42.6 $^\circ$C, 
the lasing wavelengths remain nearly unchanged.
The fluctuations can be derived with the standard deviations of 9 pm, 17 pm, and 13 pm, indicating a minor change in noise floor.      
According to the thermal response in Figure 3c, the signal-to-noise ratio (SNR) of the LC microlaser at different operating temperatures is shown in Figure 4b.   
It is found that the SNR increases from 50 at 35 $^\circ$C to 2000 at 42.6 $^\circ$C, showing a 40-fold SNR improvement in the vicinity of the LC phase transition.
Accordingly, the temperature resolution of the LC microdroplet for thermal sensing can be derived, which reaches as low as 320 \textmu K in the vicinity of phase transition.

\begin{figure*}[htbp]
\centering
\includegraphics[width=12cm]{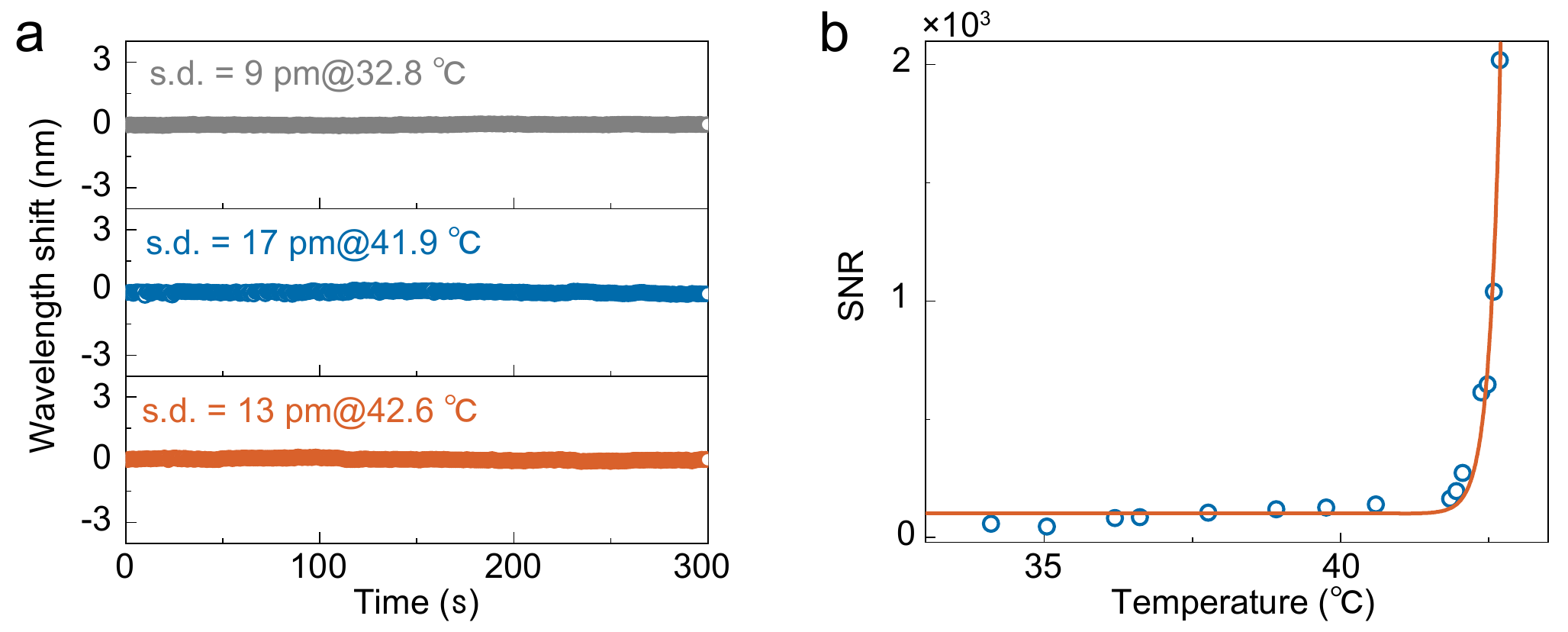}
\caption{\label{fig4}
SNR enhancement of the phase-transition microlaser for thermal sensing. (a) Stability of lasing wavelength when the LC microlaser operated at three representative temperatures. The standard deviations (s.d.) of lasing wavelengths are indicated. 
(b) SNR as a function of the operating temperature of the LC microlaser.
Circles, experimental data. Curve, fitting result. 
}
\end{figure*}

In conclusion, we have demonstrated a phase-transition microcavity laser using dye-doped LC microdroplets. 
The two phases of LC molecules and their transition process in the microcavity laser are observed experimentally, unveiling a dramatic wavelength shift of the microlaser in the vicinity of phase transition. 
This phase-transition microlaser is then applied for thermal sensing, exhibiting several new characteristics. First, the thermal sensitivity is enhanced by two-order-of-magnitude due to the phase transition, while background noises show no significant changes. Second, a temperature sensitivity of -40 nm/K and a resolution of 320 \textmu K are demonstrated, showing a distinct advantage compared to other works using the optical microcavity (Table S1).  
The stand-alone phase-transition microlaser holds great potential for high-precision temperature measurement in the life sciences such as intracellular thermal sensing.\cite{martino2019wavelength,tang2021laser,schubert2020monitoring}

\begin{acknowledgement}

This project is supported by the National Key R$\&$D Program of China (Grant 2018YFB2200401), the National Natural Science Foundation of China (Grants 11825402, 12293051, 62205007, 62105006, 11674390). X. Yang is supported by the China Postdoctoral Science Foundation (Grant 2021M700208). S.-J. Tang is supported by the China Postdoctoral Science Foundation (Grants 2021T140023 and 2020M680187).

\end{acknowledgement}


\providecommand{\latin}[1]{#1}
\makeatletter
\providecommand{\doi}
  {\begingroup\let\do\@makeother\dospecials
  \catcode`\{=1 \catcode`\}=2 \doi@aux}
\providecommand{\doi@aux}[1]{\endgroup\texttt{#1}}
\makeatother
\providecommand*\mcitethebibliography{\thebibliography}
\csname @ifundefined\endcsname{endmcitethebibliography}
  {\let\endmcitethebibliography\endthebibliography}{}

\end{document}